\documentclass[12pt]{article}
\usepackage{amsmath,amssymb}
\def\baselinestretch{1.2}
\usepackage{graphicx}
\usepackage{wrapfig}

\usepackage{hyperref}
\usepackage{cite}
\newcommand{\nb}{\bar{\nabla}}
\def\beq{\begin{eqnarray}}
\def\eeq{\end{eqnarray}}

\def\la{\langle }
\def\ra{\rangle }

\newcommand{\Tr}{\,\mathrm{Tr}\,}            

\newcommand{\Kh}{\hat{K}}

\newcommand{\CF}{{\mathcal{F}}}

\newcommand{\be}{\begin{equation}}
\newcommand{\ee}{\end{equation}}
\newcommand{\bea}{\begin{eqnarray}}
\newcommand{\eea}{\end{eqnarray}}
\newcommand{\bg}{\begin{gather}}

\newcommand{\bseq}{\begin{subequations}}
\newcommand{\eseq}{\end{subequations}}

\def\be{\begin{eqnarray}}
\def\ee{\end{eqnarray}}
\def\lb{\label}

\newcommand{\CM}{{\mathcal{M}}}

\newcommand{\p}{\partial}
\begin{document}

\title{\textbf{Boundary conformal  invariants and  the conformal anomaly in five dimensions}}

\vspace{2cm}
\author{ \textbf{  Amin Faraji Astaneh$^{a,b}$ and  Sergey N. Solodukhin$^c$ }} 
\date{}
\maketitle
\begin{center}
\hspace{-0mm}
\emph{$^a$ Department of Physics, Sharif University of Technology,\\
P.O.Box 11155-9161, Tehran, Iran}
 \end{center}
\begin{center}
\hspace{-0mm}
  \emph{$^b$ School of Particles and Accelerators,\\ Institute for Research in Fundamental Sciences (IPM),}\\
\emph{ P.O.Box 19395-5531, Tehran, Iran}
 \end{center}
\begin{center}
  \hspace{-0mm}
   \emph{ $^c$  Institut Denis Poisson UMR 7013,
  Universit\'e de Tours,}\\
  \emph{Parc de Grandmont, 37200 Tours, France} \\
\end{center}



\begin{abstract}
\noindent { In odd dimensions the integrated conformal anomaly is entirely due to  the boundary terms  \cite{Solodukhin:2015eca}.
In this paper we present a detailed analysis of the anomaly in five dimensions. We give the complete list of the boundary conformal 
invariants that exist in five dimensions. Additionally to 8  invariants known before we find a new conformal  invariant that contains the derivatives of the extrinsic curvature
along the boundary. 
Then, for a conformal scalar field satisfying either the Dirichlet or the conformal invariant Robin boundary  conditions
we use the available general results for the heat kernel coefficient $a_5$,  compute the conformal anomaly and identify the corresponding values of all boundary conformal charges.
}
\end{abstract}

\vskip 2 cm
\noindent
\rule{7.7 cm}{.5 pt}\\
\noindent 
\noindent
\noindent ~~~ {\footnotesize e-mails:\ faraji@sharif.ir ,  Sergey.Solodukhin@lmpt.univ-tours.fr}


\newpage

\section{ Introduction}
The presence of the boundaries brings the new features to the quantum field theory. One of the features is the modification of local quantities such as the heat kernel coefficients
due to the boundary terms that results in the boundary terms in the quantum effective action.
Those terms  were studied for already some time in the literature and the findings, available to the date, were summarized in a review \cite{Vassilevich:2003xt}.
The related new feature is a modification of the conformal transformations of the quantum effective action \cite{Dowker:1989ue} and of the conformal anomaly.
In the absence of the boundaries the local conformal anomaly  is non-trivial only if the dimension $d$  of the space-time is even \cite{Capper:1974ic}.
It is because  it is presented by a combination of the Euler density and the local conformal anomalies constructed form the Riemann curvature of dimension $d$.
No such invariants exist if $d$ is odd.  A related fact is that the Euler number of an odd-dimensional compact manifold is zero.

In the presence of the boundaries two new things happen. Note that in this case it is more appropriate to consider 
the integrated conformal anomaly. Then, in even dimension $d=2n$ there appear certain boundary terms in the integrated anomaly. Those boundary terms are 
constructed from the Riemann curvature of the bulk metric and the extrinsic curvature $K_{ij}$ of the boundary. Since $K_{ij}$ has dimension $1$ one can now construct terms of dimension
$2n-1$ that can be further integrated over the boundary.
These boundary terms in $d=4$ dimensions were identified in \cite{Fursaev:2015wpa}, \cite{Herzog:2015ioa}. 
The other new feature is that the anomaly in odd dimension $d=2n+1$ is non-trivial.
It is entirely  due to the boundary terms \cite{Solodukhin:2015eca} that can be of two types: the topological Euler number of the boundary $E_{2n}$  and the local conformal invariants 
constructed from the Riemann tensor and the extrinsic curvature $K_{ij}$ and their derivatives. In dimension $d=3$ there is only one such invariant so that the anomaly
is determined by two conformal charges: one for the topological term $E_2$ and the other for the conformal invariant. In \cite{Solodukhin:2015eca} these charges were determined for the 
conformal scalar both for the Dirichlet  and  the conformal Robin boundary conditions and in \cite{Fursaev:2016inw} for the Dirac fermions.

The primary goal of the present note is to advance the analysis to the next odd dimension $d=5$.  Some preliminary list of the conformal invariants in this dimension
was given in \cite{Solodukhin:2015eca}. This  list, as it was clear already in the time of writing the paper  \cite{Solodukhin:2015eca}, was rather incomplete since it was anticipated that the other, unknown at that time,  invariants 
may exist.
The essential missing element was  one or more invariants that could be constructed from the derivatives of the extrinsic curvature along the boundary  and that would not vanish if
the bulk 5d spacetime were flat.  The differential invariant presented in   \cite{Solodukhin:2015eca}  does not have this last property since, as we show this in the present paper,  it   happens to be  not an independent 
invariant as it reduces to a combination of the other invariants constructed from the various contractions of the Weyl tensor
and the extrinsic curvature tensor. 
One of the main results of the present paper is that we have now found this new invariant,  see eq.(\ref{I8}),  that contains derivatives of the extrinsic curvature tensor and that was absent in \cite{Solodukhin:2015eca}.
With the new invariant the list of the conformal boundary invariants in five dimensions is now complete.

In a wider context the present paper is a part of the on-going study of the various manifestations of the boundaries, defects and  interfaces in the conformal field theories, \cite{Herzog:2017xha} -\cite{Seminara:2018pmr}.
It should be noted that contrary to the situation with the conformal invariants in the bulk of space-time, where all such invariants are classified, no such a classification theorem exists in general for the
boundary conformal invariants.
Discussing the preliminary work we should note the earlier results in mathematical literature \cite{Glaros:2015pfa} where the conformal invariant boundary structures 
were studied in various dimensions. 
The other preliminary work relevant to our present study is the calculation of the heat kernel coefficient $a_5$ for the various boundary conditions and the different elliptic operators
that was performed in a series of papers \cite{Branson:1999jz},  \cite{Branson:1995cm},  \cite{Kirsten:1997qd} that was summarized in \cite{Kirsten:2000xc}.
It is our other principle goal in the paper to use these earlier results, compute the conformal anomaly in the case of a conformal scalar both for the Dirichlet and conformal Robin boundary conditions
and express the anomaly in terms of the complete set of the invariants thus determining the exact values of the boundary conformal charges.

It should be noted that our findings can be also  relevant to the study of the logarithmic terms in the entanglement entropy of a conformal field theory in  $d=6$.
Indeed, the entangling surface then has dimension $4$ and the logarithmic terms are due to the possible conformal invariants constructed from the curvature of spacetime and the 
extrinsic curvature of the surface. The previous study of  the logarithmic terms  within the holographic paradigm includes \cite{Safdi:2012sn}. In the paper we make contact with this
previous study.

The holographic aspect of the conformal anomaly in five dimensions in the presence of the boundaries is interesting. We, however, leave it for future work.

\section{Boundary conformal invariants in $d=5$}
\subsection{ Notations and conformal transformations}
We denote by $n^\mu$ the components of the out-ward normal vector  to the boundary $\partial {\cal M}_5$, $n^\mu n_\mu=1$. The metric induced on the boundary is $\gamma_{\mu\nu}=g_{\mu\nu}-n_\mu n_\nu$.
In this and the subsequent sections we use the notations that the latin indices $i\, , \, j\, , \, k\, \dots$ should be understood as projections along the boundary, they take values $1\, , \,  2\, , \, 3\, , \, 4$.
The index projected on the normal direction is denoted by $n$. With these notations for any tensor $X_{\mu\alpha\beta}$ one has that $X_{nij}=n^\mu\gamma_i^{\ \alpha}\gamma_j^{\ \beta}X_{\mu\alpha\beta}$.
The extrinsic curvature is defined as $K_{ij}=\gamma_{i}^{\ \mu}\gamma_{j}^{\ \nu}\nabla_{(\mu}n_{\nu )}$, where $\nabla_\mu$ is the covariant derivative defined with respect to the 5d  metric $g_{\mu\nu}$.
The covariant derivative defined with respect to the intrinsic metric $\gamma_{ij}$ is denoted by $\bar{\nabla}_i$ and the respective curvature by $\bar{R}$, $\bar{R}_{ij}$ and $\bar{R}_{ijkl}$.
The relations between the intrinsic curvature of the boundary and the curvature in the 5d space-time are given by the Gauss-Codazzi identities presented in  Appendix A. 

Under the infinitesimal  conformal transformations $\delta g_{\mu\nu}=2\sigma g_{\mu\nu}$, $\delta n_{\mu}=\sigma n_{\mu}$ the Weyl tensor transforms as $\delta W_{\alpha\mu\beta\nu}=2\sigma W_{\alpha\mu\beta\nu}$.
The extrinsic curvature transforms as follows
\be
\delta{K}_{ij}=\sigma K_{ij}+\gamma_{ij}\nabla_n\sigma \, , \ \ \ \delta K=-\sigma K+4\nabla_n\sigma\, , \ \ \ \delta\hat{K}_{ij}=\sigma\hat{K}_{ij}\, ,
\lb{2}
\ee
where $\nabla_n=n^\mu\nabla_\mu$ and $\hat{K}_{ij}=K_{ij}-\frac{1}{4}\gamma_{ij}K$ is  the trace free part of the extrinsic curvature.
The basic conformal tensors are, thus, the bulk Weyl tensor $W_{\alpha\mu\beta\nu}$ and the trace free extrinsic curvature of the boundary $\hat{K}_{ij}$. The intrinsic Weyl tensor of the boundary metric is expressed in terms of
the $5$-dimensional  Weyl tensor and the extrinsic curvature by means of the Gauss-Codazzi relations.

\subsection{Conformal anomaly}

The integrated conformal anomaly in five dimensions can be presented in the form,
\be
\int_{{\cal M}_5}\la T_{\mu\nu}\ra g^{\mu\nu}=\frac{1}{5760 (4\pi)^2}(aE_4+\sum_{k=1}^8 c_k I_k)\, ,
\lb{1}
\ee
where $E_4$ is the 4d Euler density integrated over the boundary $\partial{\cal M}_5$, so that $\chi[\partial {\cal M}_5]=\frac{1}{32\pi^2}E_4$
is the Euler number of the boundary, and $I_k\, , \, k=1,\dots , 8$ are the conformal invariants constructed from the  5-dimensional Weyl
tensor and the trace free part of the extrinsic curvature of the boundary $\hat{K}_{ij}=K_{ij}-\frac{1}{4}\gamma_{ij}K$.
We note that the Euler number of an odd-dimensional open manifold  is 1/2 of the Euler number of the boundary. 
This explains why the Euler number $E_5$  is not an independent quantity and, thus, is absent in the anomaly (\ref{1}). 
The numerical pre-factor in (\ref{1}) is chosen for the further convenience. $(a\, , \, c_1\, , \dots \, , \, c_8)$ are the boundary conformal charges to be determined below in the paper.

\subsection{Euler number of the boundary }

The Euler density of the 4-dimensional boundary  has the standard expression in terms of the intrinsic curvature of the boundary,
\be
E_4=\int_{\p\CM_5}(\bar{R}_{ikj\ell}^2-4\bar{R}_{ij}^2+\bar{R}^2)\, .
\lb{E4}
\ee
Applying the Gauss-Codazzi equations, it can be expanded in terms of the curvature tensors of the bulk manifold as well as the extrinsic curvature of the four-dimensional boundary,
\be
\begin{split}
&E_4=\int_{\p\CM_5}\Big(R^2_{ijk\ell}-4R^2_{ij}+R^2-4R_{injn}^2+8R^{ij}R_{injn}+4R^2_{nn}-4RR_{nn}+4K^{ij}K^{k\ell}R_{ikj\ell}\\&+8(K^2)^{ij}R_{ij}-8KK^{ij}R_{ij}-2\Tr K^2R+2K^2R
-8(K^2)^{ij} R_{injn}+8KK^{ij}R_{injn}\\
&+4\Tr K^2R_{nn}
-4K^2R_{nn}-6\Tr K^4+8K\Tr K^3+3(\Tr K^2)^2-6K^2\Tr K^2+K^4\Big)\, ,
\end{split}
\lb{3}
\ee
where we defined $(K^2)^{ij}=K^i_{\ k} K^{kj}$.
\bigskip

Now we list the conformal invariants which can be called algebraic. These invariants do not contain the derivatives of the extrinsic curvature.

\subsection{Invariants constructed from the extrinsic curvature tensor}
The first group of the invariants of this type is constructed from the trace free extrinsic curvature $\hat{K}_{ij}$,
\be
I_1=\int_{\p\CM_5}(\Tr\Kh^2)^2=\int_{\p\CM_5}[(\Tr K^2)^2-\frac{1}{2}K^2\Tr K^2+\frac{1}{16}K^4]\, ,
\ee
and
\be
I_2=\int_{\p\CM_5}\Tr \Kh^4=\int_{\p\CM_5}(\Tr K^4-K\Tr K^3+\frac{3}{8}K^2\Tr K^2-\frac{3}{64}K^4)\, .
\ee

\bigskip

\subsection{Invariants constructed from the Weyl tensor}
The next set of the invariants is constructed from the 5d Weyl tensor,
\be
I_3=\int_{\p\CM_5}W_{ikj\ell}^2=\int_{\p\CM_5}\left(R_{ikj\ell}^2-\frac{16}{9}R_{ij}^2+\frac{5}{18}R^2+\frac{8}{3}R^{ij}R_{injn}+\frac{4}{9}R_{nn}^2-\frac{8}{9}RR_{nn}\right)\, ,
\ee
and
\be
I_4=\int_{\p\CM_5}W_{injn}^2=\int_{\p\CM_5}\left(\frac{1}{9}R_{ij}^2-\frac{1}{36}R^2+R_{injn}^2-\frac{2}{3}R^{ij}R_{injn}-\frac{4}{9}R_{nn}^2+\frac{2}{9}RR_{nn}\right)\, .
\ee
\subsection{Invariants constructed from the Weyl tensor contracted with the extrinsic curvature tensor}
Contractions of the Weyl tensor and the trace free extrinsic curvature give us a new set of the conformal invariants,
\be
\begin{split}
I_5&=\int_{\p\CM_5}\hat{K}^{ij}\hat{K}^{k\ell}W_{ikj\ell}\\
=&\int_{\p\CM_5}\Big(K^{ij}K^{k\ell}R_{ikj\ell}+\frac{2}{3}(K^2)^{ij}R_{ij}-\frac{5}{6}KK^{ij}R_{ij}-\frac{1}{12}\Tr K^2R+\frac{1}{8}K^2R\\
&+\frac{1}{2}KK^{ij}R_{injn}-\frac{1}{6}K^2R_{nn}\Big)\, ,
\end{split}
\ee
and
\be
\begin{split}
I_6&=\int_{\p\CM_5}\hat{K}^i_k\hat{K}^{kj}W_{injn}\\
=&\int_{\p\CM_5}\Big(-\frac{1}{3}K^i_kK^{kj}R_{ij}+\frac{1}{6}KK^{ij}R_{ij}+\frac{1}{12}\Tr K^2R-\frac{1}{24}K^2R\\
&K^i_kK^{kj}R_{injn}-\frac{1}{2}KK^{ij}R_{injn}-\frac{1}{3}\Tr K^2R_{nn}+\frac{1}{6}K^2R_{nn}\Big)\, .
\end{split}
\ee

\bigskip

\subsection{Invariants with derivatives and their independence}

The invariants of this group will be expressed in terms of the derivatives of the extrinsic curvature along the boundary.
The first invariant of this kind is
\be
I_7=\int_{\p\CM_5}W_{nijk}^2\, .
\ee
Using the Gauss-Codazzi equation,  $W_{nijk}$ can be expanded in terms of the derivatives of the extrinsic curvature tensor
\be
W_{nijk}=2\nb_{[k}K_{j]i}+\frac{2}{3}\gamma_{i[j}\left(\nb_\ell K^\ell_{k]}-\nb_{k]}K\right)\, .
\ee
This identity helps us to re-write the invariant $I_7$ in the form that contains only the extrinsic curvature tensor and its derivatives along the boundary,
\be
I_7=\int_{\p\CM_5}\left(2\nb_kK_{ij}\nb^kK^{ij}-2\nb_kK^{ij}\nb_jK^k_i+\frac{4}{3}\nb_iK\nb_jK^{ij}-\frac{2}{3}\nb_iK^{ij}\nb_kK^k_j-\frac{2}{3}(\nb K)^2\right)\, ,
\ee
using \eqref{identity 4}, one re-writes this expression as follows
\be
\begin{split}
I_7&=\int_{\p\CM_5}\Big[2\nb_kK_{ij}\nb^kK^{ij}-\frac{8}{3}\nb_iK^i_j\nb_kK^{kj}+\frac{4}{3}\nb_iK\nb_jK^{ij}-\frac{2}{3}(\nb K)^2\\
&-2K^{ij}K^{k\ell}R_{ikj\ell}-2(K^2)^{ij} R_{injn}+2(K^2)^{ij} R_{ij}+2K\Tr K^3-2(\Tr K^2)^2\Big]\, .
\end{split}   
\ee

\bigskip

It is instructive to analyse the relation of this invariant to another conformal invariant that is written in terms of the conformal invariant operator acting on a tensor of rank two.
The following term is introduced in \cite{Solodukhin:2015eca}  and  is known to be a conformal invariant in 5 dimensions,
\be
I^{(D)}=\int_{\p\CM_5}\Tr \Kh \CF\Kh\, ,
\ee
where the differential operator
\be\label{conformal derivative}
\begin{split}
&\CF^{ij}_{k\ell}=\delta^i_{(k}\delta^j_{\ell)}\bar{\Box}-\frac{4}{3}\bar{\nabla}^{(i}\bar{\nabla}_{(k}\delta^{j)}_{\ell)}-\frac{8}{6}\bar{R}^i\,_{(k}\,^j\,_{\ell)}-\frac{2}{3}\bar{R}^{(i}_{(k}\delta^{j)}_{\ell)}+\frac{1}{6}\bar{R}\delta^{i}_{(k}\delta^j_{\ell)}\, 
\end{split}
\ee
acts on a tensor with conformal weight $+1$. It should be noted that there is only one such differential operator that respects the conformal symmetry  \cite{Gusynin:1986kz},  \cite{Achour:2013afa}. The explicit form of this invariant reads
\be
\begin{split}
\lb{ID}
&I^{(D)}=\int_{\p\CM_5}\Tr \Kh \CF\Kh=\int_{\p\CM_4}\Kh^{k\ell} \CF^{ij}_{k\ell}\Kh_{ij}\\
&=\int_{\p\CM_5}\Big[ K^{ij}\bar{\Box}K_{ij}-\frac{1}{3}K\bar{\Box}K+\frac{1}{3}K^{ij}\bar{\nabla}_i\bar{\nabla}_jK +\frac{1}{3}K\bar{\nabla}_i\bar{\nabla}_jK^{ij}-\frac{4}{3}K^{ij}\bar{\nabla}_j\bar{\nabla}_kK^k_i\\
&+2(K^2)^{ij}R_{injn}-KK^{ij}R_{injn}-2(K^2)^{ij}R_{ij}+KK^{ij}R_{ij}-\frac{1}{3}\Tr K^2R_{nn}+\frac{1}{3}K^2R_{nn}\\
&+\frac{1}{6}\Tr K^2R-\frac{1}{6}K^2R+2\Tr K^4-3K\Tr K^3-\frac{1}{6}(\Tr K^2)^2+\frac{4}{3}K^2\Tr K^2-\frac{1}{6}K^4\Big]\, .
\end{split}
\ee
It can be shown using the Gauss-Codazzi relations that in flat bulk spacetime  ($R_{\mu\nu\alpha\beta}=0$) all derivatives of the extrinsic curvature  in (\ref{ID})  are mutually cancelled.

In what follows it will be shown that this term is not an independent invariant. To see this, let us start with the observation that
\be
I^{(D)}=-\frac{1}{2}I_7-\int_{\p\CM_5}\hat{K}^{ij}\hat{K}^{k\ell}\bar{W}_{ikj\ell}
\ee
Note also that using the Gauss-Codazzi relations  one has that
\be
\begin{split}
&\int_{\p\CM_5}\hat{K}^{ij}\hat{K}^{k\ell}\bar{W}_{ikj\ell}=I_5+\int_{\p\CM_5}\Big[-(R_{injn}-\frac{1}{3}R_{ij})((K^2)^{ij} -\frac{1}{2}KK^{ij})\\
&+\frac{1}{3}(R_{nn}-\frac{1}{4}R)(\Tr K^2-\frac{1}{2}K^2)-2\Tr K^4+2K\Tr K^3+\frac{7}{6}(\Tr K^2)^2-\frac{4}{3}K^2\Tr K^2+\frac{1}{6}K^4\Big]\, .
\end{split}
\ee 

The integral in the right hand side of the above equation can be written in terms of $I_3$, $I_4$ and $I_6$. So finally we conclude that
\be
I^{(D)}=-\frac{1}{2}I_7-I_5+I_6-\frac{7}{6}I_3+2I_4\, .
\ee
This indicates that it is not an independent invariant and, thus,  it has to be excluded from the list.
Note that all derivative terms in $I^{(D)}$ come just from $I_7$, so it is natural that all derivative terms disappear in the flat spacetime case, where $W_{nijk}=0$.

\subsection{Invariants with derivatives: new invariant}

To construct the last invariant, we start with the normal derivative of the Weyl tensor, contracted with the normal vector and the traceless extrinsic curvature tensor
\be
I_8^{(1)}=\int_{\p\CM_5}\Kh^{ij}\nabla_n W_{injn}\, ,
\ee
where by $\nabla_n W_{injn}$ we mean $n^\rho n^\mu n^\nu \nabla_\rho W_{i\mu j\nu}$\, .\\
We can show that
\be\label{CTI81}
\delta I_8^{(1)}=-2\int_{\p\CM_5}\left(\Kh^{ij}W_{injn}\p_n\sigma+\Kh^{ij}W_{nijk}\nb^k\sigma\right)\, .
\ee
The first term in the conformal variation can be removed if we simply add 
\be
I_8^{(2)}=\frac{1}{2}\int_{\p\CM_5}K\Kh^{ij}W_{injn}\, .
\ee
Let us focus now on the second term in \eqref{CTI81}. We can construct an integral with the same transformation
\be
I_8^{(3)}=\int_{\CM_5}\left(\frac{1}{9}\nb_i\Kh^i_j\nb_k\Kh^{kj}-(\Kh^2)^{ij}\bar{S}_{ij}+\frac{1}{2}\Tr \Kh^2\bar{S}^i_i\right)\, ,
\ee
where
\be
\bar{S}_{ij}=\frac{1}{2}(\bar{R}_{ij}-\frac{1}{6}\bar{R}\gamma_{ij})\, 
\ee
is the  4d Schouten tensor computed with respect to the intrinsic boundary metric $\gamma_{ij}$.
Under the conformal transformations one has
\be
\begin{split}
&\delta(\nb_i\Kh^i_j\nb_k\Kh^{kj})=-4\sigma \nb_i\Kh^i_j\nb_k\Kh^{kj}+6\Kh^{ij}\nb_k\Kh^k_j\nb_i\sigma\, ,\\
&\delta \bar{S}_{ij}=-\nb_i\nb_j\sigma\, , \ \ \delta \bar{S}^i_i=-2\sigma \bar{S}^i_i-\bar{\Box}\sigma\, .
\end{split}
\ee
Therefore one finds that
\be
\delta I_8^{(3)}=\int_{\p\CM_5}\left[\frac{2}{3}\Kh^{ij}\nb_k\Kh^k_j\nb_i\sigma+(\Kh^2)^{ij}(\nb_i\nb_j\sigma-\frac{1}{2}\gamma_{ij}\bar{\Box}\sigma)\right]\, .
\ee
Interestingly, we may recast the above expression as follows,
\be
\delta I_8^{(3)}=\int_{\p\CM_5}\Kh^{ij}W_{nijk}\nb^k\sigma\, .
\ee
This is precisely what  we need to cancel the second term in the conformal transformation  of $I^{(1)}_{(8)}$.
So we can now construct the following conformal  invariant, $I_8=I_8^{(1)}+I_8^{(2)}+2I_8^{(3)}$,
\be
\lb{I8}
I_8=\int_{\CM_5}\left(\Kh^{ij}\nabla_n W_{injn}+\frac{1}{2}K\Kh^{ij}W_{injn}+\frac{2}{9}\nb_i\Kh^i_j\nb_k\Kh^{kj}-2(\Kh^2)^{ij}\bar{S}_{ij}+\Tr \Kh^2\bar{S}^i_i\right)\, .
\ee
Using the Gauss-Codazzi equations   and the differential relations (\ref{identity 3}), (\ref{identity 2}) and (\ref{identity 4}) 
the new invariant $I_8$ can be rewritten as follows,
\be
\lb{I8-2}
\begin{split}
I_8&=\int_{\p\CM_5}\Big[\frac{2}{3}K^{ij}\nabla_nR_{injn}-\frac{1}{12}K\nabla_nR+\frac{2}{3}K^{ij}K^{k\ell}R_{ikj\ell}-K^i_kK^{jk}R_{ij}\\
&+\frac{1}{3}\Tr K^2 R-\frac{5}{48}K^2R+\frac{5}{3}K^i_kK^{kj}R_{injn}-\frac{1}{3}KK^{ij}R_{injn}-\Tr K^2R_{nn}+\frac{11}{24}K^2R_{nn}\\
&+\Tr K^4-\frac{11}{6}K\Tr K^3+\frac{47}{48}K^2\Tr K^2-\frac{7}{48}K^4\\
&-\frac{1}{3}\nb_kK_{ij}\nb^kK^{ij}+\frac{8}{9}\nb_iK^i_j\nb_kK^{kj}-\frac{7}{9}\nb_iK^{ij}\nb_jK+\frac{25}{72}(\nb K)^2\Big]\, .
\end{split}
\ee
To the best of our knowledge the  invariant $I_8$ is a new invariant that was  not  available in the literature before. 
It has been for a long time an important  missing element
in the discussion of the conformal anomalies in five dimensions. Having said that we should notice that in some particular case this invariant reduces to the one that already
appeared in the literature on the holographic computation of the entanglement entropy in $d=6$.
Indeed, provided the  5d manifold is flat and  using the Gauss-Codazzi equations, one has that
\be
R_{nijk}=0\rightarrow \nb_kK_{ij}=\nb_jK_{ik}\, .
\ee
In this case $I_8$ takes a simpler form,
\be
I_8^{flat}=\int_{\p\CM_5}\left[\frac{1}{8}(\nb K)^2+\Tr K^4-\frac{3}{2}K\Tr K^3-\frac{1}{3}(\Tr K^2)^2+\frac{47}{48}K^2\Tr K^2-\frac{7}{48}K^4\right]\, .
\ee
Interestingly, in this form, it is related to invariant $T_3$ found by Safdi \cite{Safdi:2012sn} in a holographic calculation of the entanglement entropy in a 6 dimensional flat manifold,
\be
I_8^{flat}=\frac{1}{8}(T_3+\frac{10}{3}I_1-4I_2)\, .
\ee
It should be noted that no conformal invariant form of $T_3$ was given  in  \cite{Safdi:2012sn}.
Thus, our result  (\ref{I8}), (\ref{I8-2}), among other things,  offers a conformal invariant form, valid in a curved spacetime, for this holographic calculation.

\section{Conformal anomaly for a scalar field}
The free conformal scalar field in five dimensions is described by a conformal Laplace operator,
\be
D=-(\nabla^2+E)\, ,  \ \ E=-\frac{3}{16}R\, .
\lb{D}
\ee
In the space-time with boundaries it should be supplemented by a boundary condition. There are two possible boundary conditions consistent with the conformal symmetry:
the Dirichlet condition and the conformal Robin condition,
\be
\lb{DR}
\begin{split}
&\text{Dirichlet b.c.}\, : \, \phi\vert_{\p\CM_5}=0\, ,\\
&\text{Robin b.c.}\, :  \, (\p_n-S)\phi\vert_{\p\CM_5}=0\, ,\ \ S=-\frac{3}{8}K\, .
\end{split}
\ee
The important object that encodes the main information about the quantum field theory is the heat kernel $K(D,s)=e^{-sD}$ and its small $s$ expansion,
\be
\Tr K(D,s)=\sum_{p=0}a_p (D) \, s^{(p-5)/2}\, , \ \  s\rightarrow 0
\lb{K}
\ee
where $a_p(D)$ are the heat kernel coefficients that are represented by integrals over the manifold and its boundary  of the local invariants constructed from
the curvature of the bulk metric, the extrinsic curvature of the boundary and the quantities $E$ and $S$. 

The integrated conformal anomaly  in dimension $d=5$ is determined by the coefficient $a_5$,
\be
\int_{{\cal M}_5}\la T_{\mu\nu}\ra g^{\mu\nu}=a_5\, .
\lb{anomaly}
\ee
As was explained above, there is no a bulk term in $a_5$ and the entire contribution comes only from the boundary $\p\CM_5$.
The exact form of the coefficient $a_5$ is available in the literature for a rather general elliptic operator with the boundary condition being a mixture of the
Dirichlet and the Robin conditions, see \cite{Branson:1999jz},  \cite{Branson:1995cm},  \cite{Kirsten:1997qd}, \cite{Kirsten:2000xc}.
In what follows we use the general form of $a_5$  presented in  \cite{Branson:1999jz}.

For the conformal operator $D$ and the Dirichlet boundary condition we find that
\be
\begin{split}
a_5^{(D)}&=-\frac{1}{5760(4\pi)^2}\int_{\p\CM_5}\Big(8R_{\mu\alpha\nu\beta}^2-8R_{\mu\nu}^2+\frac{5}{16}R^2-10R_{injn}^2+16R^{ij}R_{injn}-17R_{nn}^2+\frac{5}{2}RR_{nn}\\ 
&+48\Box R-45\bar{\Box}R-\frac{111}{2}\nabla_n^2R+24\bar{\Box}R_{nn}+15\nabla_n^2R_{nn}-\frac{339}{8}K\nabla_nR\\ 
&+32K^{ij}K^{k\ell}R_{ikj\ell}+16K^i_kK^{kj}R_{ij}+14KK^{ij}R_{ij}+\frac{25}{8}\Tr K^2R-\frac{35}{16}K^2R\\
&-\frac{47}{2}K^i_kK^{kj}R_{injn}+\frac{49}{4}KK^{ij}R_{injn}-\frac{215}{8}\Tr K^2 R_{nn}-\frac{215}{16}K^2R_{nn}\\
&-\frac{327}{8}\Tr K^4+\frac{17}{2}K\Tr K^3+\frac{777}{32}(\Tr K^2)^2-\frac{141}{32}K^2\Tr K^2-\frac{65}{128}K^4\\
&+\frac{355}{8}\nb_kK_{ij}\nb^kK^{ij}-\frac{11}{4}\nb_iK^i_j\nb_kK^{jk}-\frac{29}{4}\nb_kK^j_i\nb_jK^{ik}+58\nb_iK^{ij}\nb_jK-30K\nb_i\nb_jK^{ij}\\
&-\frac{75}{2}K^{jk}\nb_k\nb_iK^i_j+\frac{285}{4}K^{ij}\nb_i\nb_jK+54K^{ij}\bar{\Box} K_{ij}+6K\bar{\Box}K-\frac{413}{16}\nb_iK\nb^iK\Big)\, ,
\end{split}
\ee
while for the conformal Robin boundary condition we have that
\be
\begin{split}
a_5^{(R)}&=\frac{1}{5760(4\pi)^2}\int_{\p\CM_5}\Big(8R_{\mu\alpha\nu\beta}^2-8R_{\mu\nu}^2+\frac{5}{16}R^2-10R_{injn}^2+16R^{ij}R_{injn}-17R_{nn}^2+\frac{5}{2}RR_{nn}\\ 
&+48\Box R-45\bar{\Box}R-\frac{111}{2}\nabla_n^2R+24\bar{\Box}R_{nn}+15\nabla_n^2R_{nn}-30K^{ij}\nabla_n R_{injn}-\frac{309}{8}K\nabla_nR\\ 
&+32K^{ij}K^{k\ell}R_{ikj\ell}+16K^i_kK^{kj}R_{ij}+\frac{43}{2}KK^{ij}R_{ij}-\frac{5}{8}\Tr K^2R-\frac{5}{4}K^2R\\
&+\frac{133}{2}K^i_kK^{kj}R_{injn}-\frac{161}{4}KK^{ij}R_{injn}-\frac{275}{8}\Tr K^2 R_{nn}-\frac{185}{16}K^2R_{nn}\\
&+\frac{231}{8}\Tr K^4-\frac{125}{4}K\Tr K^3+\frac{375}{32}(\Tr K^2)^2+\frac{147}{32}K^2\Tr K^2-\frac{37}{64}K^4\\
&+\frac{535}{8}\nb_kK_{ij}\nb^kK^{ij}+\frac{49}{4}\nb_iK^i_j\nb_kK^{jk}+\frac{151}{4}\nb_kK^j_i\nb_jK^{ik}+58\nb_iK^{ij}\nb_jK-\frac{75}{2}K\nb_i\nb_jK^{ij}\\
&-\frac{15}{2}K^{jk}\nb_k\nb_iK^i_j+\frac{315}{4}\nb_i\nb_jK+114K^{ij}\bar{\Box} K_{ij}-\frac{3}{2}K\bar{\Box}K-\frac{503}{16}\nb_iK\nb^iK\Big)\, .
\end{split}
\ee
We shall insert here
\be
R^2_{\mu\alpha\nu\beta}=R_{ikj\ell}^2+4R_{nijk}^2+4R_{injn}^2\, , \ \ R_{\mu\nu}^2=R_{ij}^2+2R_{in}^2+R_{nn}^2\, .
\ee
The other important point is that the general expression for the coefficient $a_5$  given in  \cite{Branson:1999jz} contains some sort of redundancy since the differential relation (\ref{identity 5}), which together with (\ref{identity 2})  generalises the Gauss-Codazzi identities,
was not taken into account. In fact, a similar redundancy is present in the coefficient $a_4$ given in  \cite{Vassilevich:2003xt},  where the relation\footnote{The relation (\ref{identity 2}) has appeared earlier in \cite{McAvity:1990we} and  \cite{Fursaev:2015wpa}.} (\ref{identity 2}) was not taken into account.  
Since these relations   contain the higher order normal derivatives one might worry whether one  would have  to specify an extension for the normal vector outside the boundary to make these relations valid.
We, however, have checked that the relations  (\ref{identity 1}) - (\ref{identity 5})  do not depend on  the way the normal vector is extended. 
Thus,  using the Gauss-Codazzi equations  and the identities (\ref{identity 1}), (\ref{identity 5}) we finally arrive at  the expression
\be
\begin{split}
a_5^{(D)}&=\frac{1}{5760(4\pi)^2}\int_{\p\CM_5}\Big(-8R_{ikj\ell}^2+8R_{ij}^2-\frac{5}{16}R^2-22R_{injn}^2-R^{ij}R_{injn}+10R_{nn}^2-\frac{5}{2}RR_{nn}\\ 
&-15K^{ij}\nabla_n R_{injn}+\frac{15}{8}K\nabla_nR\\ 
&+\frac{277}{4}K^{ij}K^{k\ell}R_{ikj\ell}-\frac{289}{4}K^i_kK^{kj}R_{ij}+KK^{ij}R_{ij}-\frac{25}{8}\Tr K^2R+\frac{35}{16}K^2R\\
&+\frac{499}{4}K^i_kK^{kj}R_{injn}-\frac{109}{4}KK^{ij}R_{injn}-\frac{25}{8}\Tr K^2 R_{nn}-\frac{25}{16}K^2R_{nn}\\
&+\frac{327}{8}\Tr K^4-\frac{379}{4}K\Tr K^3+\frac{1983}{32}(\Tr K^2)^2+\frac{141}{32}K^2\Tr K^2+\frac{65}{128}K^4\\
&-\frac{555}{8}\nb_kK_{ij}\nb^kK^{ij}+\frac{165}{2}\nb_iK^i_j\nb_kK^{jk}-\frac{135}{4}\nb_iK^{ij}\nb_jK+\frac{285}{16}\nb_iK\nb^iK\Big)\, 
\end{split}
\ee
for the conformal scalar operator with the Dirichlet boundary condition  and
\be
\begin{split}
a_5^{(R)}&=\frac{1}{5760(4\pi)^2}\int_{\p\CM_5}\Big(8R_{ikj\ell}^2-8R_{ij}^2+\frac{5}{16}R^2+22R_{injn}^2+R^{ij}R_{injn}-10R_{nn}^2+\frac{5}{2}RR_{nn}\\ 
&-15K^{ij}\nabla_n R_{injn}+\frac{15}{8}K\nabla_nR\\ 
&-\frac{97}{4}K^{ij}K^{k\ell}R_{ikj\ell}+\frac{109}{4}K^i_kK^{kj}R_{ij}+\frac{13}{2}KK^{ij}R_{ij}-\frac{5}{8}\Tr K^2R-\frac{5}{4}K^2R\\
&+\frac{41}{4}K^i_kK^{kj}R_{injn}-\frac{101}{4}KK^{ij}R_{injn}-\frac{35}{8}\Tr K^2 R_{nn}+\frac{55}{16}K^2R_{nn}\\
&+\frac{231}{8}\Tr K^4+10K\Tr K^3-\frac{945}{32}(\Tr K^2)^2+\frac{147}{32}K^2\Tr K^2-\frac{37}{64}K^4\\
&+\frac{255}{8}\nb_kK_{ij}\nb^kK^{ij}-\frac{105}{2}\nb_iK^i_j\nb_kK^{jk}+\frac{135}{4}\nb_iK^{ij}\nb_jK-\frac{255}{16}\nb_iK\nb^iK\Big)\, 
\end{split}
\ee
for the conformal scalar operator with the conformal Robin boundary condition (\ref{DR}).

Following the earlier works  \cite{Branson:1999jz} - \cite{Kirsten:2000xc}, we adopted in this paper  the convention that
the components of the normal vector $n^\mu$ are always outside the covariant derivatives, for instance $\nabla_n R_{nn}\equiv n^\alpha n^\mu n^\nu \nabla_\alpha R_{\mu\nu}$,
$\nabla_n R_{ninj}\equiv n^\alpha n^\mu n^\nu \nabla_\alpha R_{\mu i \nu j}$ etc. This guarantees that these terms do not depend  on  how  the components of the normal vector are extended outside the boundary.
It should be noted that the elliptic operators in question do not require any such extension so that the respective heat kernel coefficients contain only terms that are independent of the way the vectors normal to the boundary 
are extended to the nearest vicinity of the boundary. This, in particular, explains why in the heat kernel coefficients there do not appear  any terms that contain the normal derivative of the extrinsic curvature, $\nabla_n K_{ij}$,
since for these terms one would need  to specify the extension of  the normal vector  outside the boundary\footnote{We thank D. Vassilevich for correspondence on this point.}.

The expressions for $a_5$, obtained above,  should be now compared with the general form (\ref{1}) of the anomaly decomposed over the possible conformal invariants.
We have checked that the decomposition is indeed possible and unique both for $a_5^{(D)}$ and $a_5^{(R)}$ so that the list of conformal invariants
$E_4\, , \, I_1\, , \dots \, , \, I_8$ is indeed complete. The details of the analysis are given in Appendix B. 
The decomposition allows us to determine the values of the boundary conformal charges in the case of a conformal scalar field.
For each boundary condition, one arrives at 27 equations on 9 charges. That solution of this highly overdetermined system of algebraic equations exists is a nice check on our formulas.
The result is presented in  Table 1. The  values found for the Euler charge $a$ agree with the earlier computations   \cite{Rodriguez-Gomez:2017aca}, \cite{Nishioka:2021uef}, \cite{Wang:2021mdq}.
\begin{table}\label{table}
\begin{center}
\renewcommand{\baselinestretch}{2}
\medskip
\caption{ Conformal charges for Dirichlet and Robin boundary conditions}
\bigskip
 \begin{tabular}{ | c | c | c | }
    \hline
  \text{Conformal charges} & \text{Dirichlet b.c.} & \text{Robin b.c.} \\
  \hline
  $a$ & $\frac{17}{8}$ & $-\frac{17}{8}$ \\
  \hline
$c_1$ & $-\frac{681}{32}$ & $\frac{39}{32}$ \\
  \hline
$c_2$ & $\frac{609}{8}$ & $\frac{309}{8}$ \\
  \hline
 $c_3$ & $-\frac{81}{8}$ & $\frac{81}{8}$ \\
  \hline
$c_4$ & $-\frac{27}{2}$ & $\frac{27}{2}$ \\
  \hline
$c_5$ & $-\frac{9}{8}$ & $\frac{189}{8}$ \\
  \hline
$c_6$ & $\frac{819}{8}$ & $\frac{441}{8}$ \\
  \hline
$c_7$ & $-\frac{615}{16}$ & $\frac{195}{16}$ \\
  \hline
$c_8$ & $-\frac{45}{2}$ & $-\frac{45}{2}$ \\
  \hline
\end{tabular}
\renewcommand{\baselinestretch}{1}
\end{center}
\end{table}

\section{Conclusions}
In this note we have presented an exhaustive discussion of the boundary conformal invariants in five dimensions. In particular, and this is one of our main results, we have found a new conformal invariant
$I_8$ that contains the derivatives of the extrinsic curvature tensor along the boundary. In a flat space limit this invariant is related  to the one found holographically by Safdi.
We note that the invariants discussed in this paper do not depend on  the way the normal vector is extended outside the boundary.
Only invariants of this type may appear in the heat kernel of an elliptic operator that knows nothing about such an extension.
We, however, note here for the record that there  exists a family of the conformal tensors and the respective conformal invariants that can be defined provided an extension of the normal vector is given.
These invariants have not been   discussed in this paper.   

We use the available in the literature general results for the heat kernel coefficient $a_5$  and compute the integrated conformal anomaly for a conformal scalar
satisfying  either the Dirichlet or the conformal Robin boundary conditions. The anomaly is then uniquely decomposed over the set of conformal invariants we have found.
We then  compute the respective conformal charges. This is our second main result. It would be interesting to extend our results and compute the conformal anomaly and the respective boundary  charges for 
the conformal fields of higher spin. Although we do not anticipate any principle difficulties we do not present this analysis here leaving it to the future.

Among other possible applications, our results will be useful in classifying the possible conformal invariants that appear in the logarithmic terms of entanglement entropy of a $d=6$ conformal
field theory. 
The holographic derivation  of the conformal anomaly  in  five dimensions is yet another interesting subject to explore. We leave these directions 
for   future work.

\section*{Acknowledgements} 
We thank  Cl\'ement Berthiere for collaboration at the beginning  of this project. We are grateful to Dmitri Vassilevich for a useful correspondence.
This project was started when both of us were visiting the Theory Division at CERN in the spring of 2018. We thank CERN for  hospitality.

\newpage
\appendix
\section{Gauss-Codazzi relations}
\setcounter{equation}0
\numberwithin{equation}{section}
The Gauss-Codazzi identities give us relations between the bulk curvature $\cal R$, the intrinsic curvature on the boundary $\bar{\cal R}$ and the extrinsic curvature
$K_{ij}$,
\be\label{GC1}
R_{ikjl}=\gamma^\mu_i \gamma^\alpha_k \gamma^\nu_j \gamma^\beta_\ell R_{\mu\alpha\nu\beta}=\bar{R}_{ikj\ell}-(K_{ij}K_{k\ell}-K_{i\ell}K_{jk})\, ,
\ee
\be\label{GC2}
R_{nijk}= \gamma^\mu_i \gamma^\nu_j \gamma^\rho_k  n^\alpha  R_{\alpha\mu\nu\rho}=(\bar{\nabla}_kK_{ij}-\bar{\nabla}_jK_{ik})\, ,
\ee
where $\gamma^\mu_i$ represents the projection operation.\\ \\
\underline{\textbf{Contracted Equations}}
\be\label{CGC1}
R_{in}=R_{ni}=(\bar{\nabla}_jK^j_i-\bar{\nabla}_iK)\, .
\ee
\be\label{CGC2}
R_{ij}=\bar{R}_{ij}+ R_{injn}+(K^2_{ij}-KK_{ij})\, .
\ee
\underline{\textbf{Doubly Contracted Equations}}
\be
R=\bar{R}+2 R_{nn}+(\Tr K^2-K^2)\, .
\ee
Thus, one finds for $G_{nn}=G_{\mu\nu}n^\mu n^\nu$,  $G_{\mu\nu}=R_{\mu\nu}-\frac{1}{2}g_{\mu\nu}R$,
\be
G_{nn}=-\frac{1}{2}\bar{R}-\frac{1}{2}(\Tr K^2-K^2)\, .
\ee
\\
\underline{\textbf{Differential Equations}} 
\be\label{identity 1}
\Box R=\bar{\Box}R+\nabla_n^2R+K\nabla_n R\, ,
\ee
\be\label{identity 2}
\nabla_nG_{nn}=\nabla_n(R_{nn}-\frac{1}{2}R)=K^{ij}R_{ij}-KR_{nn}-\bar{\nabla}_i\bar{\nabla}_jK^{ij}+\bar{\Box}K\, ,
\ee
\be\label{identity 3}
\begin{split}
\nabla_nR_{ij}&=\nabla_nR_{injn}-2K^{k\ell}R_{ijk\ell}-K^k_iR_{jnkn}-K^k_jR_{inkn}+KR_{injn}+K_{ij}R_{nn}\\
&-\Tr K^2K_{ij}+KK_{ik}K^k_j+\nb_i\nb_kK^k_j+\nb_j\nb_kK^k_i-\bar{\Box}K_{ij}-\nb_i\nb_jK\, ,
\end{split}
\ee
\be\label{identity 4}
\begin{split}
\nb_k K_{ij}\nb^j K^{ik}=\nb_iK^i_j\nb_kK^{kj}+K^{ij}K^{k\ell}R_{ikj\ell}-K^{ik}K^j_kR_{ij}\\
+K^{ik}K^j_kR_{injn}-K\Tr K^3+(\Tr K^2)^2+T.D.\, ,
\end{split}
\ee
\be\label{identity 5}
\begin{split}
 \nabla_n^2G_{nn}&=-R^{ij}R_{injn}+R_{nn}^2
-K^i_kK^{kj}R_{ij}+\Tr K^2R_{nn}\\
&+K^{ij}\nabla_nR_{ij}-K\nabla_n R_{nn}-\nb_iK^{ij}\nb_jK+(\nb K)^2+T.D.\, ,
\end{split}
\ee
where we defined $\nabla_n G_{nn}=n^\alpha n^\mu n^\nu \nabla_\alpha G_{\mu\nu}$ and  $ \nabla_n^2 G_{nn}=n^\alpha n^\beta n^\mu n^\nu \nabla_\alpha\nabla_\beta G_{\mu\nu}$.
\section{Algebraic equations for the conformal charges}
Matching the coefficients of the various terms  in $a_5^{(D,R)}$  with the corresponding factors in the decomposition over the basis of the conformal invariants one arrives at the following algebraic equations for the conformal charges:
\begin{equation*}
\scriptsize{
\begin{split}
&\Tr K^4\ : \ -6a+c_2+c_8 = (\frac{327}{8},\frac{231}{8})\, , \\
&K\Tr K^3\ : \ 8a-c_2+2c_7-\frac{11}{6}c_8=(-\frac{379}{4},10)\, ,\\
&(\Tr K^2)^2\ : \ 3a+c_1-2c_7=(\frac{1983}{32},-\frac{945}{32})\, ,\\
&K^2\Tr K^2\ : \ -6a-\frac{1}{2}c_1+\frac{3}{8}c_2+\frac{47}{48}c_8=(\frac{141}{32},\frac{147}{32})\, , \\
&K^4\ : \ a+\frac{1}{16}c_1-\frac{3}{64}c_2-\frac{7}{48}c_8=(\frac{65}{128},-\frac{37}{64})\, ,\\
&R_{ikj\ell}^2\ : \ a+c_3 = (-8,8)\, , \\
&R_{ij}^2\ : \ -4a-\frac{16}{9}c_3+\frac{1}{9}c_4=(8,-8)\, ,\\
&R^2\ : \ a+\frac{5}{18}c_3-\frac{1}{36}c_4=(-\frac{5}{16},\frac{5}{16})\, ,\\
&R_{injn}^2\ : \ -4a+c_4=(-22,22)\, , \\
&R^{ij}R_{injn}\ : \ 8a+\frac{8}{3}c_3-\frac{2}{3}c_4=(-1,1)\, ,\\
&R_{nn}^2\ : \ 4a+\frac{4}{9}c_3-\frac{4}{9}c_4=(10,-10)\, ,\\
&RR_{nn}\ : \ -4a-\frac{8}{9}c_3+\frac{2}{9}c_4=(-\frac{5}{2},\frac{5}{2})\, , \\
&K^{ij}K^{k\ell}R_{ikj\ell}\ : \ 4a+c_5-2c_7+\frac{2}{3}c_8 =(\frac{277}{4},-\frac{97}{4})\, , \\
&K^i_kK^{kj}R_{ij}\ : \ 8a+\frac{2}{3}c_5-\frac{1}{3}c_6+2c_7-c_8 =(-\frac{289}{4},\frac{109}{4})\, , \\
&KK^{ij}R_{ij}\ : \ -8a-\frac{5}{6}c_5+\frac{1}{6}c_6 =(1,\frac{13}{2})\, , \\
&\Tr K^2R\ : \ -2a-\frac{1}{12}c_5+\frac{1}{12}c_6+\frac{1}{3}c_8 =(-\frac{25}{8},-\frac{5}{8})\, , \\
&K^2R\ : \ 2a+\frac{1}{8}c_5-\frac{1}{24}c_6-\frac{5}{48}c_8 =(\frac{35}{16},-\frac{5}{4})\, , \\
&K^i_kK^{kj}R_{injn}\ : \ -8a+c_6-2c_7+\frac{5}{3}c_8=(\frac{499}{4},\frac{41}{4})\, , \\
&KK^{ij}R_{injn}\ : \ 8a+\frac{1}{2}c_5-\frac{1}{2}c_6-\frac{1}{3}c_8 =(-\frac{109}{4},-\frac{101}{4})\, , \\
&\Tr K^2R_{nn}\ : \ 4a-\frac{1}{3}c_6-c_8=(-\frac{25}{8},-\frac{35}{8})\, , \\
&K^2R_{nn}\ : \ -4a-\frac{1}{6}c_5+\frac{1}{6}c_6+\frac{11}{24}c_8 =(-\frac{25}{16},\frac{55}{16})\, , \\
&\nb_kK_{ij}\nb^kK^{ij}\ : \ 2c_7-\frac{1}{3}c_8=(-\frac{555}{8},\frac{255}{8})\, , \\
&\nb_iK^i_j\nb_kK^{kj}\ : \ -\frac{8}{3}c_7+\frac{8}{9}c_8=(\frac{165}{2},-\frac{105}{2})\, , \\
&\nb_iK_{ij}\nb_jK\ : \ \frac{4}{3}c_7-\frac{7}{9}c_8=(-\frac{135}{4},\frac{135}{4})\, , \\
&(\nb K)^2\ : \ -\frac{2}{3}c_7+\frac{25}{72}c_8=(\frac{285}{16},-\frac{255}{16})\, ,\\
&K^{ij}\nabla_n R_{injn}\ : \ \frac{2}{3}c_8=(-15,-15)\, ,\\
&K\nabla_n R\ : \ -\frac{1}{12}c_8=(\frac{15}{8},\frac{15}{8})\, .
\end{split}
}
\end{equation*}
The numbers in the parentheses correspond to the (Dirichlet, Robin) boundary conditions, respectively. These 27 equations can be solved consistently to determine uniquely  9 conformal charges. The results have been listed in Table 1.

\end{document}